\documentclass[twocolumn, aps, prb, showpacs, preprintnumbers, a4paper]{revtex4-1}
\usepackage[T1]{fontenc}
\usepackage{mathrsfs}
\usepackage{amssymb}
\usepackage{amsmath}
\usepackage{graphicx}
\usepackage{braket}
\usepackage{physics}
\usepackage{bm}
\usepackage{bbold}
\usepackage[utf8]{inputenc}
\usepackage{color}
\usepackage[colorlinks=true,citecolor=blue,urlcolor=blue]{hyperref}

\begin{document}

\title{Proximity spin-orbit coupling in graphene on alloyed transition metal dichalcogenides}
\author{Zahra Khatibi}
\affiliation{School of Physics, Trinity College Dublin, Dublin 2, Ireland}
\author{Stephen R. Power}
\email{stephen.r.power@dcu.ie}
\affiliation{School of Physical Sciences, Dublin City University, Ireland}
\affiliation{School of Physics, Trinity College Dublin, Dublin 2, Ireland}

\date{\today}

\begin{abstract}
    The negligible intrinsic spin-orbit coupling (SOC) in graphene can be enhanced by proximity effects in stacked heterostructures of graphene and transition metal dichalcogenides (TMDCs).
    The composition of the TMDC layer plays a key role in determining the nature and strength of the resultant SOC induced in the graphene layer.
    Here, we study the evolution of the proximity--induced SOC as the TMDC layer is deliberately defected. 
    Alloyed ${\rm G/W_{\chi}Mo_{1-\chi}Se_2}$ heterostructures with diverse compositions ($\chi$) and defect distributions are simulated using 
    density functional theory.
    Comparison with continuum and tight-binding models allows both local and global signatures of the metal-atom alloying to be clarified.
    Our findings show that, despite some dramatic perturbation of local parameters for individual defects, the low--energy spin and electronic behaviour follow a simple effective medium model which depends only on the composition ratio of the metallic species in the TMDC layer. 
    Furthermore, we demonstrate that the topological state of such alloyed systems can be feasibly tuned by controlling this ratio.
\end{abstract}
\maketitle

\section{Introduction}
Spintronics exploits the spin degree of freedom of an electron to store and transfer information in a similar manner to how the carrier charge is used in conventional electronic devices.
Spin--orbit coupling (SOC) is an essential ingredient in many spintronic devices, as it couples the spin and charge degree of freedom, allowing a spin current to be created and controlled by electrical means \cite{van2016spin,soumyanarayanan2016emergent,galceran2021control}. 
Furthermore, it is responsible for exotic spin Hall effects, generating a transverse spin flow when an electrical current passes through a material \cite{Sinova2015spin,Ferreira2014Extrinsic,Huang2016Direct,garcia2017spin,phong2017effective,Island2019Spin,Serlin2020Intrinsic,Vila2021valley} and visa versa \cite{Inoue2003Diffuse,garello2013symmetry}.
Graphene can serve as an excellent spin channel in spin--logic devices, thanks to its long spin diffusion length and lifetime \cite{Pesin2012graphene,han2014graphene}.
However, the intrinsic SOC in graphene is too weak to be feasibly exploited to generate or manipulate spin currents \cite{Min2006intrinsic,Gmitra2009band}. 
There are several proposals for the enhancement of SOC in graphene, including hydrogenation \cite{Ferreira2014Extrinsic} or the introduction of impurities \cite{Lundeberg2013defect,Pachoud2014Scattering,soriano2015spin}, such as vacancies or heavy metal adatoms \cite{pi2010manipulation,weeks2011engineering,ma2012strong}.
These defects, however, can form local magnetic moments \cite{Yazyev2007defect,Hector2016atomic} that may lead to increased scattering and suppress spin transport \cite{Kochan2014spin,soriano2015spin,Nguyen2021large}.

\begin{figure}[tpb]
	\center
	\includegraphics[width=\linewidth]{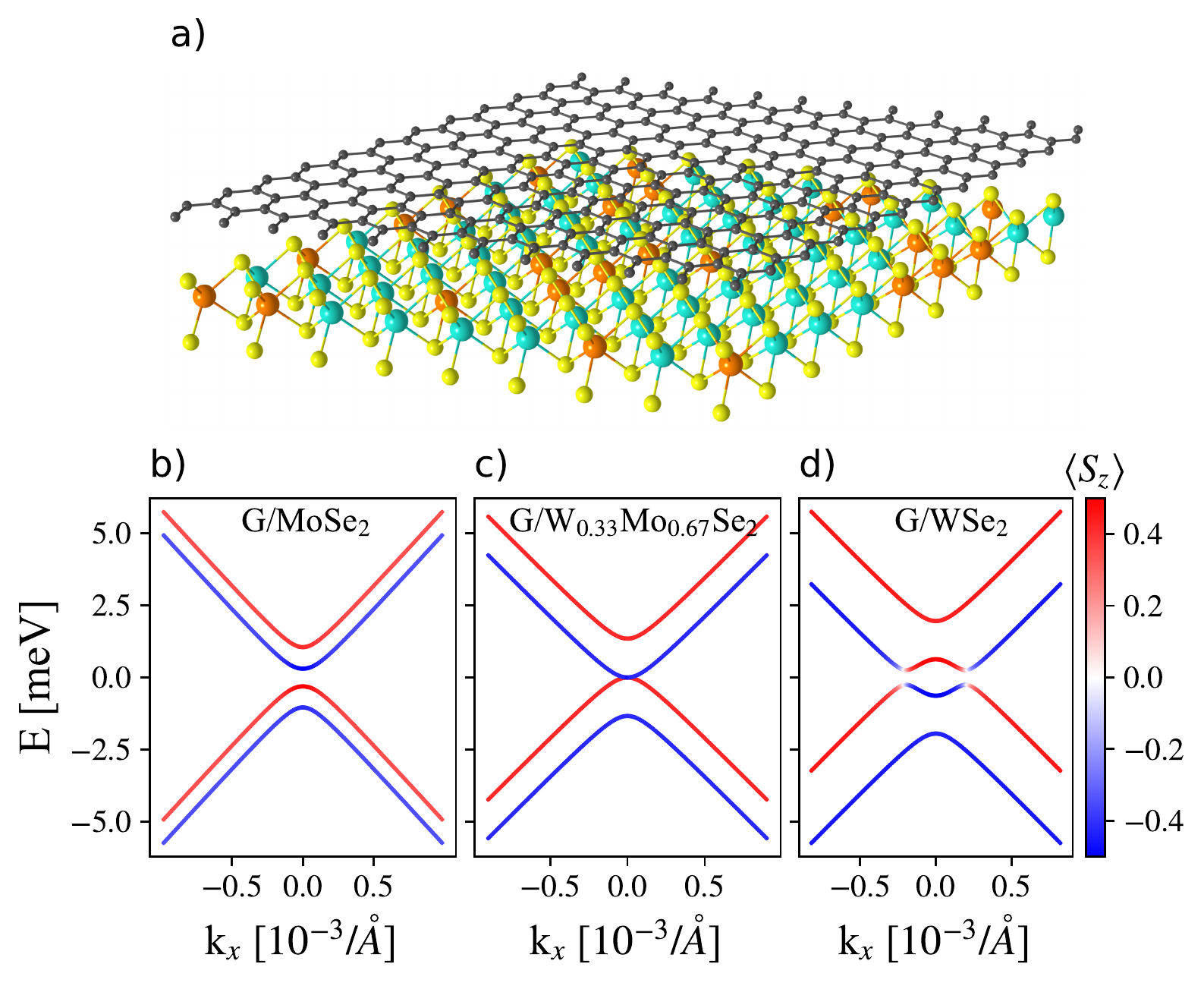}
	\caption{(a) Schematic representation of a composite G/W$_{\chi}$Mo$_{1-\chi}$Se$_2$ heterostructure. The top layer is graphene and the bottom layer is the alloyed TMDC layer. The orange, cyan, and yellow spheres in the lower layer indicate molybdenum, tungsten and selenium atoms, respectively.
	(b-d) The proximity-induced band structure and topology transition for different TMDC layers. 
	The W concentrations in the G/W$_{\chi}$Mo$_{1-\chi}$Se$_2$ systems are (b) $\chi=0$, (c) $\chi=0.33$, and (d) $\chi=1$. The band colors indicate the spin projection along the transverse axis ($z$).  \label{graphene/TMDC_lattice}}
\end{figure}

A promising alternative is to use substrates with strong SOC.
Transition metal dichalcogenides (TMDCs) are particularly interesting as they induce a SOC with a strong valley-dependence via proximity effects \cite{avsar2014spin,wang2015strong,Gmitra2016trivial,Alsharari2016mass,Yang_2016,yan2016two,Wang2016origin,cummings2017giant,dankert2017electrical,ghiasi2017large,Tobias2017Magnetotransport,Yang2017strong,FRIEDMAN201818,Garcia2018Spin,Wakamura2018strong,Singh2018Structural,Alsharari2018Topological,benitez2018strongly,Offidani2018Microscopic,arora2020superconductivity,Kumar2021zero,sierra2021van,Tiwari2021Electric,Josep2021Electrical}. 
Based on Hanle experiments and WAL measurements, spin dephasing in graphene/TMDCs is seen to be governed by the D’yakonov–-Perel mechanism near the Dirac point \cite{Yang2017strong,Tobias2017Magnetotransport,FRIEDMAN201818} and is dominated by valley--Zeeman (VZ) SOC \cite{Wang2016origin,cummings2017giant,Tiwari2021Electric}.
The proximity--induced SOC, and associated imprinted spin--valley locking, enable experimentally verified spin-charge conversion and anisotropic spin relaxation effects that are absent in pristine graphene \cite{cummings2017giant,benitez2018strongly,Offidani2018Microscopic,Josep2021Electrical}.
The nature of the proximity-induced SOC depends on the specific TMDC substrate \cite{Gmitra2016trivial}. 
The proximity effects can open an optical band gap due to a sublattice-asymmetric mass term, as in  Fig.~\ref{graphene/TMDC_lattice}(b) for a stacked G/MoSe$_2$ heterostructure \cite{Alsharari2016mass,Gmitra2016trivial}. 
A similar mass term is associated with topological valley currents in graphene--hexagonal boron nitride superlattices\cite{Gorbachev2014}, although there is some debate about how this mechanism relates to experimental observations\cite{zhu2017edge, aktor2021, Roche_2022}.
In other cases, the proximity to TMDCs lead to VZ--driven inverted bands \cite{kane2005z}, as in Fig.~\ref{graphene/TMDC_lattice}(d) for a G/WSe$_2$ heterostructure \cite{Gmitra2016trivial}.
Despite similarities with topological insulators, these systems have a trivial $\mathbb{Z}_2$ index\cite{Frank2018Protected, Alsharari2018Topological}. 
However, the valley-projected electronic spectrum of such states yield a nonzero Chern number that enables the formation of topologically-protected pseudohelical modes in finite-sized ribbons \cite{Frank2018Protected}.
The direct gap and band inversion regimes are connected through a semimetallic band gap closure which is accessible by adjustments to the effective potential difference between the layers \cite{Alsharari2016mass}.
An analytical study on twisted graphene/TMDCs heterostructures with the focuse on emerging topologically non-trivial states by enhancing otherwise weak Kane--Mele SOC shows that the SOC transfer is robust to twists between the layers \cite{Alsharari2018Topological}.
However, the disappearance of SOC at specific twist angles has also been reported \cite{David2019induced}.
The induced SOC can also be controlled using strain, where the vertical compression of graphene/WSe$_2$ using hydrostatic pressure leads to an enhancement of the VZ SOC \cite{fulop2021boosting}.

Recent advances in the growth of defect--free TMDC lateral heterostructures using water--assisted CVD techniques \cite{sahoo2018one,li2020general,Zhu2020one} suggest a new approach to tuning SOC effects in stacked graphene/TMDC heterostructures. 
These techniques enable a precise control over the sequential formation of TMDC heterojunctions by switching between the reactive gas environments \cite{sahoo2018one}.
Furthermore, it has been shown that the electronic properties of TMDC lateral heterostructures can be modulated by controlling the metal atom composition \cite{nugera2022bandgap}.
These findings, and the fact that individual TMDC family members have unique SOC signature in graphene/TMDC heterostructures, motivate the study of how the metal atom composition of the TMDC layer affects the nature and strength of the resultant induced SOC.

In this work, we investigate how proximity effects from a range of different composite TMDC layers alter the electronic and spin-orbit characteristics of the graphene layer.
We consider mixed 
G/W$_{\chi}$Mo$_{1-\chi}$Se$_2$ 
heterostructures with diverse composition ratios ($\chi$) and distributions in the TMDC layer. 
An example of such a heterostructure is shown schematically in Fig. \ref{graphene/TMDC_lattice}(a). 
Utilizing first--principals calculations and a continuum model approach, we evaluate disorder--induced local SOC signatures in the spin- and electronic behaviour of these systems.
Our density functional theory (DFT) calculations allow us to probe the microscopic origins of local SOC changes, whereas Dirac Hamiltonian and tight--binding (TB) model give further insights into the relative strength of different spin-orbit mechanisms and the emergent behaviour expected at device scales. 
We develop an effective medium model, based on the metal-atom composition ratio, which accurately captures the low-energy spin- and electronic responses in large-scale alloyed structures. 
This model predicts a semimetallic phase transition as the composition ratio is varied, as shown in Fig. \ref{graphene/TMDC_lattice}(b)-(d).  
Since G/MoSe$_2$ and G/WSe$_2$ heterostructures individually maintain robust direct and inverted band states \cite{Gmitra2016trivial,Frank2018Protected}, this demonstrates that the topological state of alloyed systems can be feasibly tuned via controlling the composition ratio of metallic element. 

This article is organized as follows: In Section \ref{geo}, we present the geometry and the structural details of the graphene/TMDC lattice considered throughout the study. 
In Section \ref{eff_model}, an effective medium model and its associated SOC parameters for alloyed systems are introduced, and a comparison is made with the DFT calculations.  
Next, a tight--binding model is used in Section \ref{TB_correct} to examine local perturbations induced in heterostructures. 
We then investigate in Section \ref{larg_comp_syst} how local corrections to the TB model affect the predictions of the effective model and, in particular,  the topological status of large composite systems.
Finally we summarize our analysis and findings in section \ref{conclusions}.

\section{Geometrical Considerations}\label{geo}
We consider a supercell structure of 4$\times$4 graphene and 3$\times$3 TMDC unit cells, an example of which is presented in Fig.~\ref{SOC_param}(a). Utilising the DFT \textsc{Quantum} ESPRESSO package \cite{Giannozzi_2009}, we calculate the electronic properties of heterostructures with MoSe$_2$, WSe$_2$ and various W$_{\chi}$Mo$_{1-\chi}$Se$_2$ alloys as the lower-lying TMDC layer.
Beginning with pristine MoSe$_2$ and WSe$_2$ bottom layers, we first split the residual lattice mismatch equally between the graphene and TMDC layers and then relax the full supercell. 
Table.~\ref{Table_structal_param} contains the structural information about each of the relaxed structures. 
During the structural optimization, we find that the full supercell in each case expands in order to reduce the forces and stress on the stiffer graphene layer. 
In a realistic system, the graphene and TMDC layers are highly unlikely to form a commensurate stacked structure but rather an incommensurate one which leaves the graphene layer unstrained. 
Thus it is important that the graphene layer is not significantly strained in our electronic structure calculations.

\begin{table}[bp]
\begin{center}
\begin{tabular}{ccccccc}
\hline\hline
& TMDC &$a$ & $a_{\rm G}$ & strain & corrugation & interlayer dis. \\
& & [\AA~] & [\AA~] & [\%] & [pm] & [\AA~] \\
\hline
&MoSe$_2$ & 3.295  & 2.471 &  +0.45 &  1.88 & 3.385 \\
&WSe$_2$ & 3.297& 2.473 & +0.52  &  2.64 & 3.382 \\
\hline
\hline
\end{tabular}
\end{center}
\caption{{\small Structural properties of graphene/TMDC heterostructures after a full DFT structural optimization. $a(a_{\rm G})$ is the lattice constant in the TMDC (graphene) layer. The next column shows the in-plane strain in the graphene layer, compared to a freestanding layer with 2.46\AA~lattice constant. The last two columns indicate the corrugation of the graphene surface and interlayer distance between the adjacent layers, respectively.}} \label{Table_structal_param}
\end{table}

The MoSe$_2$ and WSe$_2$ relaxed slabs maintain an identical structure with almost equal lattice constants, ie. 3.295 and 3.297 \AA.
Since there are no important qualitative differences between the properties of MoSe$_2$ systems calculated using both geometries, we take the relaxed WSe$_2$ structure as the fixed geometry for the systems considered in the remainder of our calculations. 
This allows an easy interchange of metal atoms when we consider alloyed systems in the following sections. 
More details about the DFT methodology employed in our calculations are presented in Appendix \ref{DFT}.

\begin{table}[bpt]
	\begin{center}
		\begin{tabular}{cccccccc}
			\hline\hline
			 TMDC &  $t$ & $\Delta$ & $\lambda_{\rm R}$  & $\lambda_{\rm I}$ & $\lambda_{\rm VZ}$ & $\lambda_{\rm PIA}$  & $\Delta_{\rm PIA}$ \\
			&[eV] & [meV] &  [meV] & [$\mu$eV] & [meV] & [meV] & [meV]\\ \hline
			
			\hline
			MoSe$_2$	&	2.53	&	-0.59	&	0.29	&	-3.87	&	0.28	&	-7.14	&	-0.86	\\
			WSe$_2$	&	2.531	&	-0.52	&	0.51	&	-3.06	&	1.15	&	0.39	&	-0.47	\\ 
			\hline\hline
		\end{tabular}
	\end{center}
	\caption{{\small Calculated orbital and spin-orbit parameters of graphene in a graphene/TMDC heterostructures, found by fitting DFT results to a continuum Dirac model. $t$ is the nearest neighbor tunneling energy, $\Delta$ is the proximity--induced orbital gap, $\lambda_{\rm I}$ is the intrinsic spin-orbit coupling, $\lambda_{VZ}$ is the valley Zeeman spin--orbit coupling, $\lambda_{\rm R}$ is the Rashba SOC, and $\lambda_{\rm PIA}$ and $\Delta_{\rm PIA}$ are the pseudospin--inversion--asymmetry spin--orbit terms.}} \label{Tab:graphene/TMDC_param}
\end{table}
%
%
%
%

\section{Continuum model for alloyed systems}\label{eff_model}
The low-energy band structure and spin texture of proximitized graphene, in the vicinity of the Dirac points, are well--described using a continuum model approach \cite{Gmitra2016trivial, Kochan2017Model}.
The parameters required for this model, $\{\Lambda\}=(t, \Delta, \lambda_{\rm I}, \lambda_{\rm VZ}, \lambda_{\rm R}, \lambda_{\rm PIA}, \Delta_{\rm PIA})$, include spin-independent hopping ($t$) and mass ($\Delta$) terms, in addition to the strengths of the intrinsic (I), valley Zeeman (VZ), Rashba (R) and pseudospin--inversion--asymmetry (PIA) spin-orbit terms that arise due to interactions between the layers.
Here, the mass term is a spin--independent gap--opening term that emerges due to the net effect of local sublattice--symmetry breaking. 
The intrinsic term is a standard Kane--Mele type coupling \cite{kane2005z} which introduces a spin--independent topological gap, whereas the VZ term is sublattice--asymmetric version which rigidly shifts the Dirac cone according to the spin and valley indices.
These terms do not mix spin channels, unlike the Rashba coupling, a substrate-induced term arising due to symmetry--breaking in the direction perpendicular to graphene layer.
Finally, the two PIA terms are required to correctly capture features further from the Dirac points. 
The Hamiltonian for each of these terms is discussed in more detail in Appendix~\ref{appendixcont}.

The  parameter sets $\{\Lambda\}$ for different heterostructures can be found by fitting the electronic band structure and spin expectation values from first--principles calculations to the continuum model.
We begin by considering pristine G/MoSe$_2$ and G/WSe$_2$ heterostructures, whose DFT calculated energy spectrum and associated spin projections around the K valley are shown in Fig. \ref{graphene/TMDC_lattice}(b) and (d). 
While the energy spectrum of G/MoSe$_2$ is parabolic with a finite direct gap, we see an inverted band dispersion for G/WSe$_2$, with an avoided crossing and a flip of the spin orientation along the axes normal to the supercell plane ($s_z$).
The continuum model, with the fitted orbital and spin--orbit parameters given in Table \ref{Tab:graphene/TMDC_param}, perfectly reproduces all the electronic and spin features of full DFT calculations.
We note that both the DFT results and the extracted parameters for these pristine systems are in good agreement with those first obtained in Ref. \onlinecite{Gmitra2016trivial}.
Both G/MoSe$_2$ and G/WSe$_2$ systems show a band gap, but the topological nature of the gap in each system is different, which can be understood in terms of the interplay of mass, VZ and Rashba contributions.
The Kane--Mele SOC is significantly smaller than the other terms.
Furthermore, because of the staggered nature of intrinsic SOC in G/TMDC heterostructures, these systems are topologically trivial with a vanishing  topological order \cite{Frank2018Protected}.
While the mass term, $\Delta$, opens a optical gap, the VZ term acts to close it by introducing a spin-dependent band shift in each valley. 
If $\lambda_{\rm VZ} > \Delta$, the band gap is closed in each valley by overlapping bands with different spin orientations.
A finite $\lambda_{\rm R}$ couples the two spin channels, giving rise to band anticrossings and band inversion. 
While not a true topological insulator, this VZ-driven regime gives rise to  time--reversal symmetry protected edge states\cite{Frank2018Protected}.
The relative strength of the mass, VZ and Rashba terms is thus of vital importance in determining the overall nature of the graphene/TMDC heterostructure.

The orbital and SOC parameters depend on the composition of the TMDC layer, as shown by comparing the results for MoSe$_2$ and WSe$_2$ in Table \ref{Tab:graphene/TMDC_param}.
The choice of metal atom determines the induced SOC, whereas the chalcogen atom affects the alignment of the Dirac cone with the bulk TMDC bands~\cite{Gmitra2016trivial}.
We note that the value of the sublattice-staggered mass term, and hence the system band topology, can be sensitive to geometric details.
Studies of twisted graphene/TMDC structures suggests that this term may vanish in the incommensurate limit \cite{wang2015strong,Yang_2016,Alsharari2018Topological,Naimer2021Twist}, whereas for commensurate structures it varies with the supercell size used in DFT calculations.
The VZ term, in contrast, remains approximately constant \cite{Gmitra2016trivial, wang2015strong}.
Our choice of selenide-based structures is motivated by the robust direct and inverted band phases of the two pure systems.
G/WSe$_2$ has a VZ parameter which is significantly larger than the mass term for even the smallest supercells.
In contrast, G/WS$_2$ sits near the semimetallic phase transition and studies with slightly different setups report different topological behaviour \cite{Gmitra2016trivial, wang2015strong}.
As the G/MoSe$_2$ system has a sizeable direct gap, alloyed ${\rm G/W_{\chi}Mo_{1-\chi}Se_2}$ systems are likely to present a wide range of behaviours as we tune the composition ratio $\chi$.
In particular, there must be an intermediate set of parameters between those of G/MoSe$_2$ and G/WSe$_2$ which closes the band gap, and separates the direct and inverted band phases, as shown in Fig. \ref{graphene/TMDC_lattice}(c).

\begin{figure}[tp]
	\includegraphics[width=\linewidth]{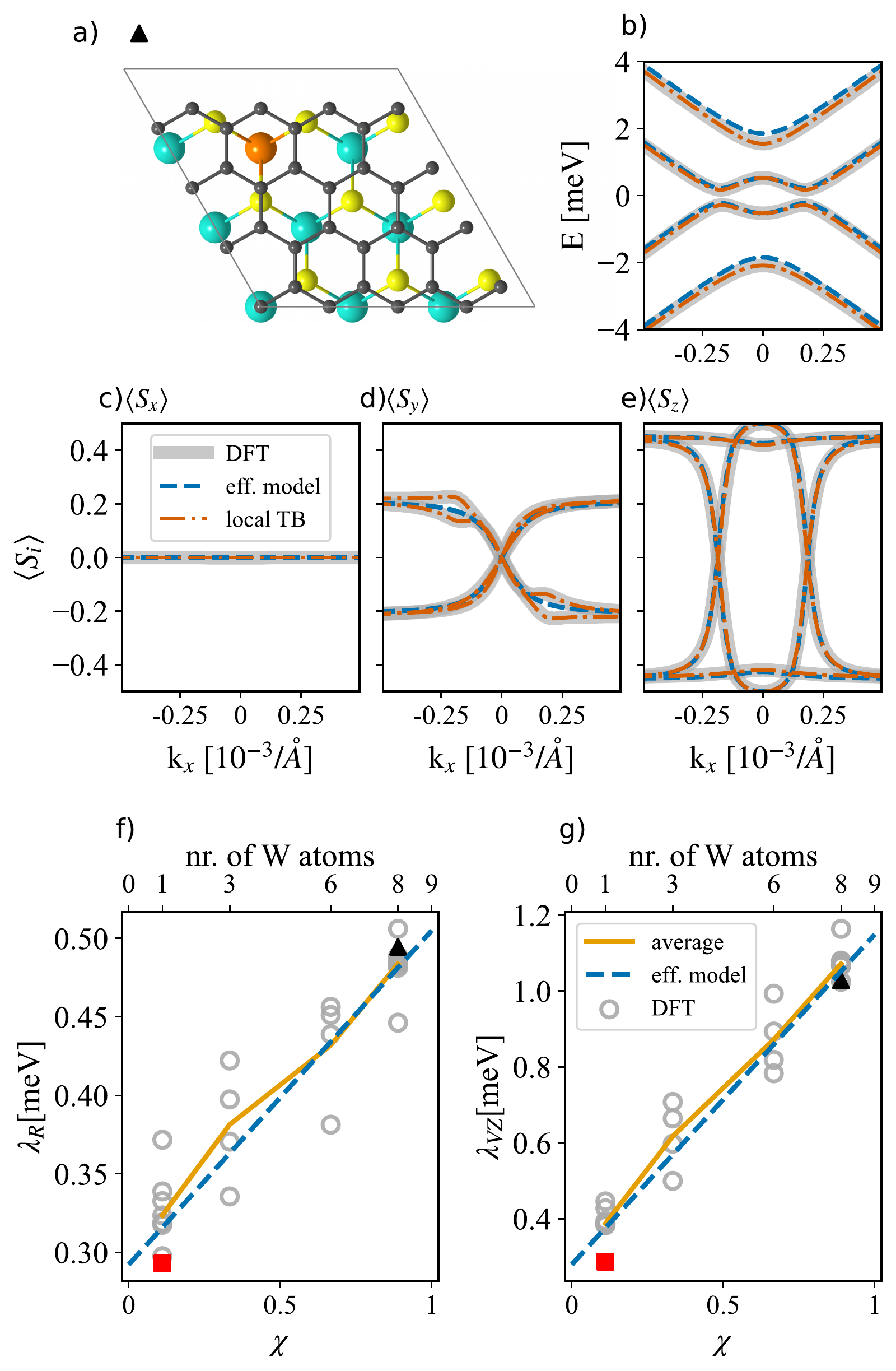}
	\caption{
		(a) The supercell of an example alloyed system. The orange sphere is an Mo atom substitutionally--doped into a G/WSe$_2$ heterostructure. 
		(b) Electronic band structure and (c--e) spin texture,  $\langle S_i\rangle$ with $i=x,y,z$, of the alloyed system shown in panel (a). 
		(f,g) The evolution of Rashba and VZ SOC parameters as the number of W atoms in the unit cell is varied. The grey circles show the SOC parameters extracted from DFT calculations for individual configurations with different impurity distributions. The orange curve shows the average of these extracted parameters over all  configurations with the same concentration. The blue curve shows the effective model result found using a weighted average of pristine system parameters.
	\label{SOC_param}}
\end{figure}

To model the evolution of the electronic and spin--orbit behaviour of alloyed systems, we first consider an effective model using a weighted average (WAVG) of the parameters extracted from the pristine systems.
The WAVG parameters in a composite system G/W$_{\chi}$Mo$_{1-\chi}$Se$_2$ are taken to be $\chi \, \{\Lambda_{\rm W}\}\ + (1-\chi) \, \{\Lambda_{\rm Mo}\} $, where $\{\Lambda_{\rm Mo (W)}\}$ is the set of parameters for the pristine Mo (W) structure.
This model assumes a uniform modulation of the model parameters as the concentration of W atoms is increased, and can be used to make predictions for arbitrary systems.
The alloyed band structure and spin-texture in Fig. \ref{graphene/TMDC_lattice}(c), which show a critical point where the gap closure occurs, were generated using this model for $\chi=0.33$.
Comparing the parameters for pristine Mo and W systems in Table \ref{Tab:graphene/TMDC_param}, we note that the most significant variation occurs for the VZ and Rashba terms.
As discussed above, it is the interplay of these terms with the more slowly-varying $\Delta$ term which will determine the band topology of the system.

To test the validity of the WAVG model, we performed DFT calculations for a range of composite systems where the TMDC layer contains a mixture of Mo and W atoms. 
We employed the same 9 metal atom unit cell, and chose configurations to ensure we sampled a diverse set of metal atom distributions relative to the carbon atoms in the graphene layer. 
For example, Fig. \ref{SOC_param}(a) is one of the nine possible structures with 8 W atoms and 1 Mo atom in the TMDC layer, and the impurity Mo atom lies directly below the centre of a hexagon in the graphene layer.
Other impurity positions are less symmetric with respect to the graphene lattice.
The thick grey curves in Fig. \ref{SOC_param} (b)--(e) present the DFT band structure and spin texture for this system, which are similar to those of a pristine G/WSe$_2$ heterostructure.
The defect--induced modulation of the mass term is weak and thus the VZ SOC still dominates the gap, leading to a band inversion in the presence of a Rashba term.
We note that, for this system, both the band structures and spin textures are in near--perfect agreement with the WAVG effective model, whose predictions are shown by the dashed blue lines in Fig. \ref{SOC_param} (b)--(e).

The band structure and spin expectation values for all configurations were then fully fitted with the effective model to find the orbital and spin-orbit parameters $\{\Lambda_i\}$ which 
best described the system.
The extracted parameters for the key Rashba and VZ terms of every composite system are shown by the symbols in Fig.~\ref{SOC_param} (f),(g), as a function of the composition ratio $\chi$.
The full set of parameters for the single-impurity configurations are given for reference in Table \ref{Tab:impurity_param} of the Appendix.
The specific system discussed above, with 8 W atoms in the unit cell, is shown by a solid black triangle in Fig.~\ref{SOC_param} (f),(g).
Compared to the concentration--dependent values predicted by the WAVG model (dashed blue line), the values for the individual configurations show some deviation, but overall follow a similar trend and tend to adopt values between the pristine limiting cases.
Furthermore, the configurational average of the individual cases, shown by the orange line, agrees almost exactly with the WAVG effective model for both Rasha and VZ terms. 
This suggests that the electronic and spintronic behaviour of large--scale alloyed systems, containing a mix of impurity locations, can be largely understood with the effective model.
In the next section, we discuss cases where local DFT results seem to disagree with the weighted--average approach.

\begin{figure}[tp]
	\center
	\includegraphics[width=.9\linewidth]{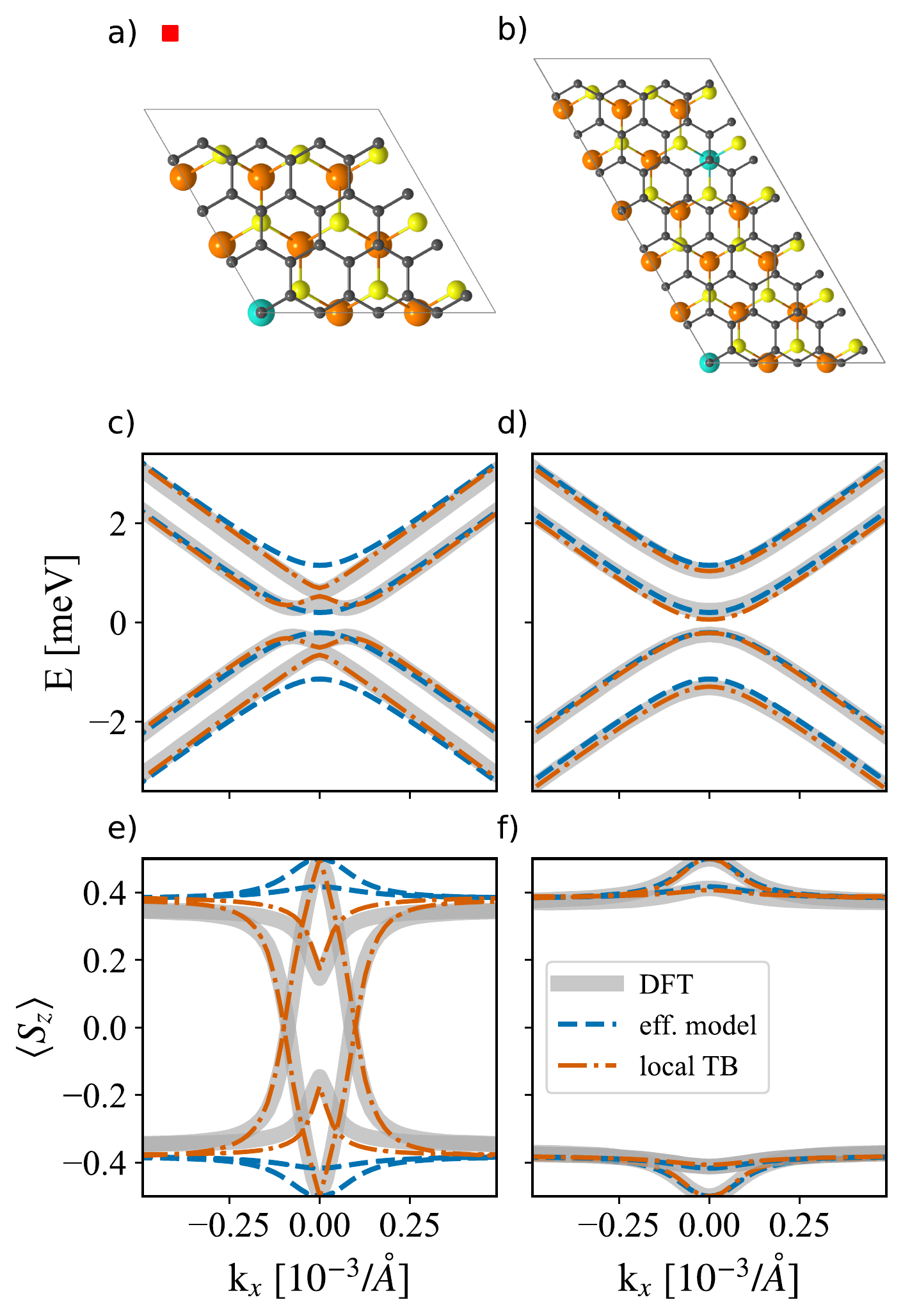}
	\caption{(a) One block and (b) two block supercells of alloyed G/MoSe$_2$ heterostructures. (c,d) The energy spectrum and (e,f) the spin expectation value along the $z$ axis for the alloyed heterostructures shown in (a,b).
	The alloyed G/MoSe$_2$ supercell shown in panel (b) is twice as large as the supercell in panel (a). 
	It contains one impurity with the same, and one with different, positioning relative to the graphene lattice, thereby reducing the  symmetry of the impurity distribution. 
 \label{1vs2_blk}}
\end{figure}

\section{Local effects in alloyed systems}\label{TB_correct}
The effective model, although largely successful in predicting the electronic behavior of alloyed systems, does not account for variations across the orbital and SOC parameters for individual configurations.
For example, the extracted Rashba and VZ parameters for the system represented by red squares in Fig.~\ref{SOC_param} (f) and (g) are both smaller than the WAVG model predicts (dashed blue line).
This system, which contains a single W atom in the TMDC unit cell, is considered in detail in the left-hand panels of Fig. \ref{1vs2_blk}.
We note that that the energy bands (panel c) and spin projection $\langle S_z \rangle$ (panel e) of the DFT calculation differ from the WAVG model prediction. 
The WAVG model predicts that, with only one W atom out of 9 metal atom sites, the system should have a direct gap, revealed by the constant--sign behaviour of $\langle S_z \rangle$.  
However,  the DFT results, show a small band inversion and a flip in the sign of $\langle S_z \rangle$ right at the Dirac point, indicative of a system which is barely to the inverted band side of the critical point separating the phases.
These discrepancies are partially explained by the continuum nature of the WAVG effective model, which assumes uniform potential and spin--orbit fields throughout the system. 
However, even for uniform MoSe$_2$ or WSe$_2$ bottom layers, this is only an approximation.
For example, the unit cells we employ in this work, such as those shown in Figs. \ref{SOC_param}(a) and \ref{1vs2_blk}(a) contains 9 metal atoms in the TMDC layer.
Each of these align slightly differently with the carbon atoms in the graphene lattice above, and will therefore induce slightly different local potential and SOC fields.
Continuum models assume that such differences quickly average out and give rise to uniform \emph{effective} proximity effects throughout the graphene layer.
However, DFT calculations are limited to small unit cells, so that this averaging is not complete, and even for pristine TMDC layers a small relative shift between the layers can give rise to slightly different band structures and SOC parameters \cite{Gmitra2016trivial}.

To determine the role that local variation of hopping, potential and spin--orbit parameters can play in alloyed systems, we now move to TB models which can account for such real--space atomistic--level perturbations.
We will use this method to add local corrections, taking into account the local metal--atom alignment near individual impurities, on top of the effective medium model which accounts for the overall metal--atom composition ratio $\chi$.
We consider the systems shown in Figs.~\ref{SOC_param}(a) and \ref{1vs2_blk}(a) as examples.
Although the energy bands and spin texture of the former, shown in Fig.~\ref{SOC_param}(b)--(e), are largely in agreement with the WAVG model, there are some subtle differences.
The DFT bands show a slight asymmetric spin--dependent shift of the valence and conduction bands, which is not captured by the effective model.
This suggests that some combination of the mass term, and intrinsic, VZ and Rashba SOC terms are locally perturbed from the WAVG values.
A splitting of the in--plane spin expectation values in Fig. \ref{SOC_param}(d) is further indicative of a local perturbation of the Rashba parameter.
Within a TB model (see Appendix \ref{app:tb}), we account for these by allowing local perturbations to onsite potentials ($\delta V_i$) and the Rashba and Kane--Mele hopping terms ($\delta \lambda_{\rm R}^{ij}, \delta \lambda_{\rm I}^{ij}$) associated with specific carbon atoms $i,j$ near the impurity metal atom. 
For this system, we vary five local fitting parameters, with the best--fit values allowing fine details from the DFT calculations to be captured, as demonstrated by the orange, dot-dashed curves in Fig. \ref{SOC_param}(b)--(e).

The DFT results for the Mo--rich heterostructure in the left--hand panels of Fig.~\ref{1vs2_blk} show that the introduction of a single W atom into the unit cell can induce changes far beyond what is predicted by the effective model. 
This suggests that certain geometrical arrangements in alloyed systems can strongly affect local electronic and spin--orbit properties.
This is particularly so for systems near the critical point, where small parameter changes can give rise to band--inversion and a sign--flip in the spin texture. 
The dramatic changes to the electronic and spin character of the system require four spin--independent parameters in the tight--binding correction, namely onsite potentials applied on the carbon site directly over the impurity W atom and also to its three nearest neighbours. 
This alters the effective mass term locally, and gives an excellent fit to the DFT results (dot--dashed orange curves), including the changes to the band topology and spin--texture.
It is important to note that, aside from local changes to the electronic and spin--orbit fields, DFT simulations can also introduce artefacts that are not necessarily representative of realistic alloyed systems.
Due to the small size of the periodic supercell, spurious symmetric replicas of the defects can effectively interact with each other\cite{Venezuela_2009}.
In addition, symmetry and periodicity in their own right can introduce band gap effects in graphene\cite{dvorak2013bandgap}.
It is therefore important to verify that any local TB corrections we make are accurately capturing local changes to the potential and spin-orbit fields, and are not trying to reproduce periodicity artefacts. 
To rule out such effects, we also perform calculations and fittings with a larger supercell, i.e. twice the size of the standard cell used so far.
To test the local parameterization of the W impurity shown in Fig.~\ref{1vs2_blk}(a), we extend the cell along one of the standard cell vectors, as shown in  Fig.~\ref{1vs2_blk}(b).
To maintain the same composition ratio $\chi$ and further reduce the symmetry of the system, we introduce a second W defect in the cell.
The DFT electronic spectrum and spin textures shown by grey curves in Fig.~\ref{1vs2_blk}(d) and (f) are in much closer agreement to the WAVG effective model. 
Performing a TB fitting near the impurity in this cell yields a much weaker perturbation to the local mass term.
This suggests that the single-cell fitting shown in Fig.~\ref{1vs2_blk}(c) and (e) was not representative of the effect this impurity will have in truly disordered alloyed layers.
We have tested the TB corrections for each of the single-impurity systems similarly, by checking that they also give an improvement beyond the effective model in larger cells.

\begin{figure}[tp]
	\center
	\includegraphics[width=\linewidth]{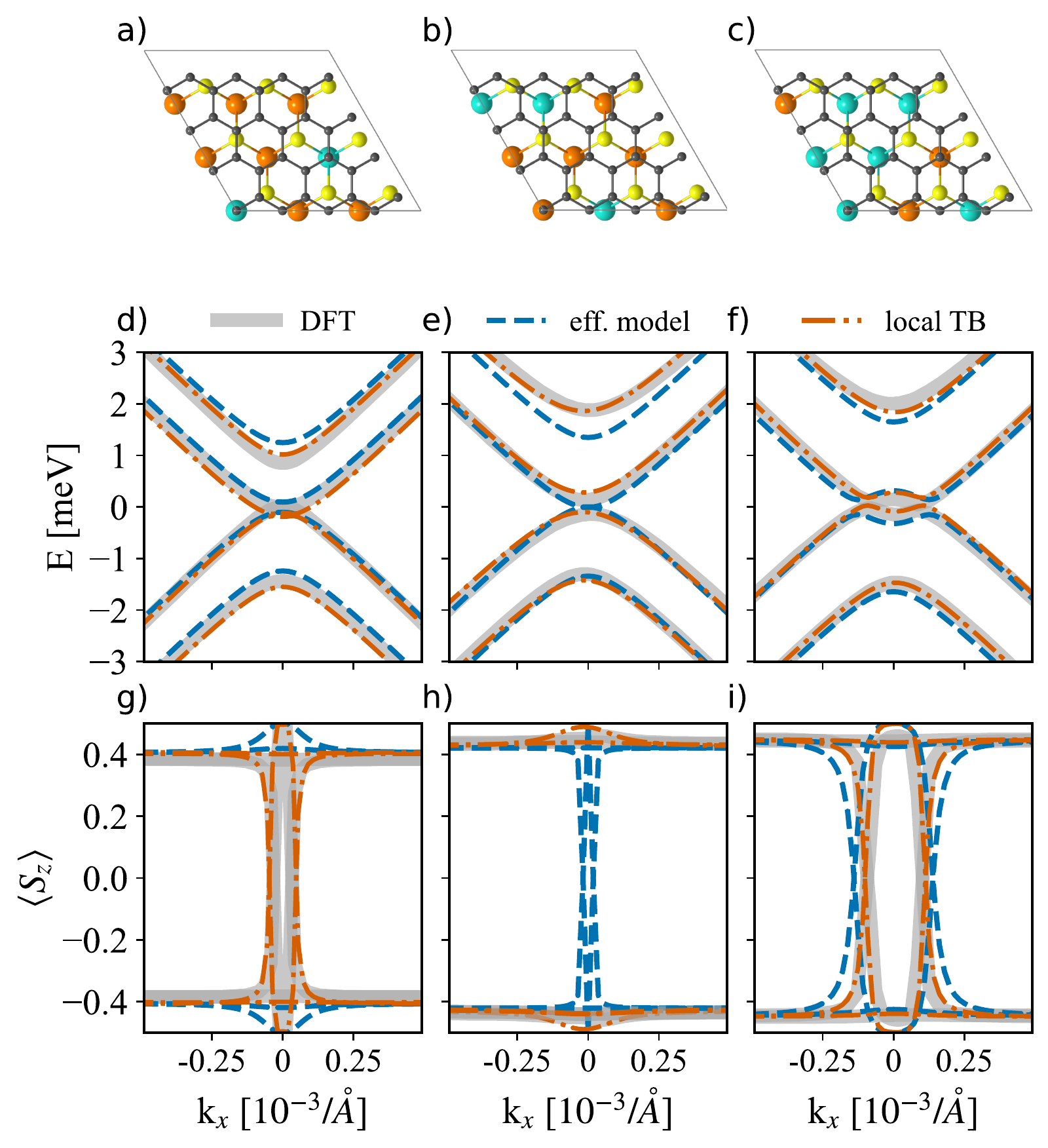}
	\caption{(a--c) The supercells of composite G/W$_{\chi}$Mo$_{1-\chi}$Se$_2$ structures with diverse metal atom composition. (d--f) Energy dispersion and (g--i) $z$ component of the spin expectation value for mixed systems shown in (a--c). The local TB results presented here are calculated by combining individual local TB perturbations from single alloyed systems with the effective model. 
	\label{combine_imp}}
\end{figure}

The single impurity systems provide us with a valuable microscopic insight to understand more complex systems, particularly where DFT results deviate from the WAVG effective model.
In Fig.~\ref{combine_imp}(a)--(c) we consider three systems with multiple metal atom impurities in the standard, smaller supercell.
Similar to the case of larger supercells, a very good agreement is seen between the DFT (grey) and effective model (blue dashed) band structures in Fig.~\ref{combine_imp}(d)--(f). .
This suggests that local effects near individual impurities can begin to cancel out when multiple impurities are present, and also that periodicity effects are reduced when the metal atom distribution is less symmetric.
More significant deviations between the DFT and effective models are seen in the spin textures in Fig.~\ref{combine_imp}(g)--(i).
This is particularly evident for the first two systems which are Mo--rich and fall on either side of the critical point between inverted (left) and normal (right) gap behaviour.
In both cases, the effective model predicts the opposite behaviour, which is most clearly seen by examining the discrepancies between the solid grey and blue dashed curves in Fig.~\ref{combine_imp}(g) and (h).
However, the DFT behaviour can be recovered by adding the combined TB corrections for each of the W impurities to the WAVG model, as shown by the dot-dashed orange curves. 
It is worth emphasising that the TB corrections here do not require any additional fitting, but instead use the individual impurity parameters extracted from single- or double supercell calculations, as discussed above. 
The excellent agreement with DFT, even for systems near the phase transition, underscores the reliability of the TB corrections, which can now be used to study larger systems beyond the scope of DFT methods.

\section{Larger TB simulations for complex systems}\label{larg_comp_syst}

\begin{figure}[tpb]
	\center
	\includegraphics[width=\linewidth]{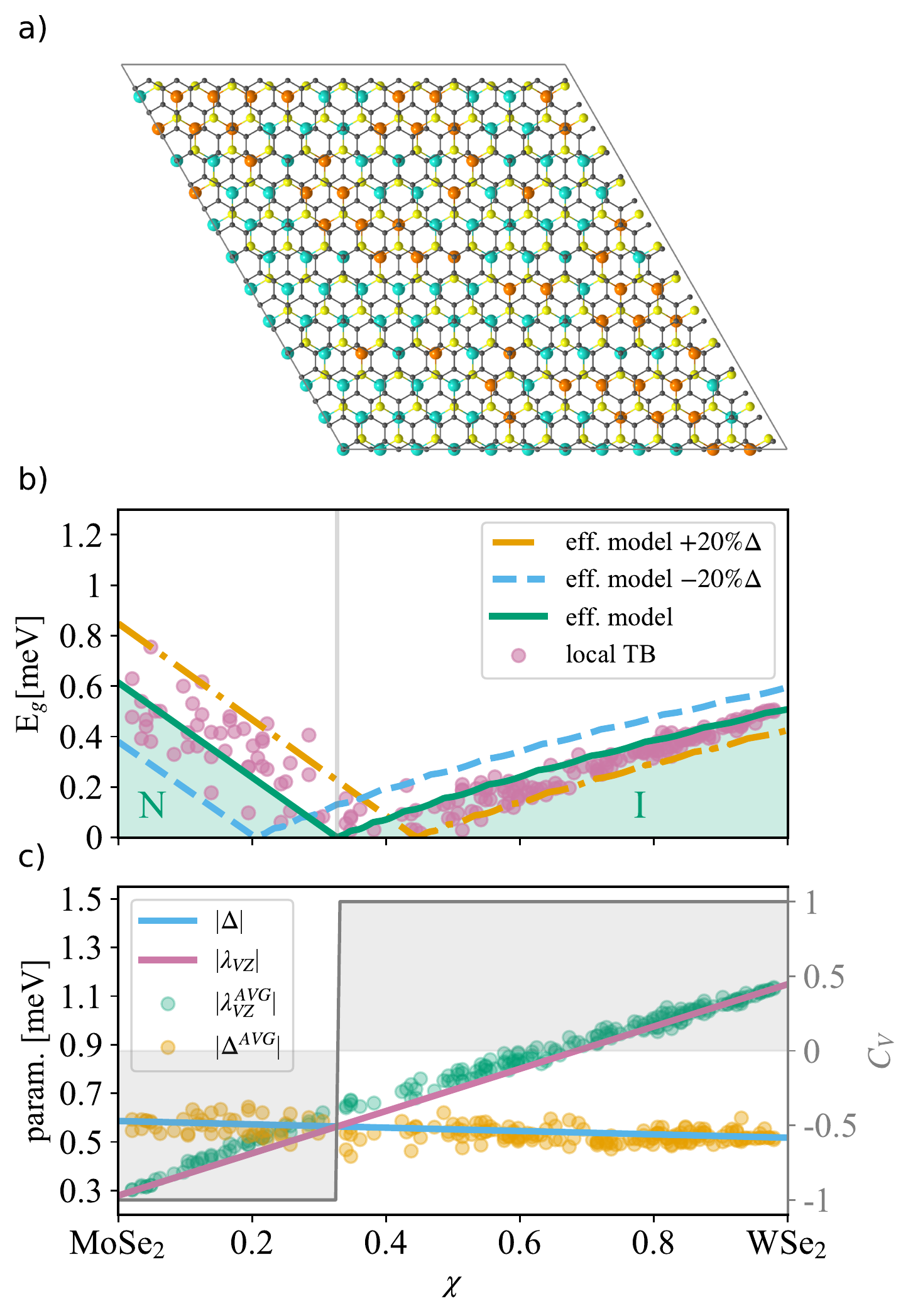}
	\caption{(a) Schematic of a composite 4$\times$4 blocks graphene/TMDC cell used in tight-binding calculations.
	(b) Energy gap, (c) mass and VZ parameters of composite systems as a function of the composition ratio $\chi$. 
	The solid green curve in (b) shows the gap predicted by the effective model as a function of the composition ratio $\chi$, which closes at $\chi=0.33$ (grey gridline).
	This is a semimetallic critical point that separates normal and inverted gap regimes.
	Purple disks show the calculated band gaps for 4$\times$4 blocks composite systems using local TB corrections
	The solid lines and symbols in (c) similarly show the effective model and supercell-averaged values of the mass and VZ parameters in these systems.
	The dashed curves in (b) show the effect of a $\pm20\%$ variation in the mass term in the effective model, which shifts the critical concentration. 
	The grey curve in (c) shows the sign change of the valley Chern index at the critical point.
	\label{gap_imp}}
\end{figure}

Our large supercell and multiple-impurity DFT calculations hint that realistic alloyed systems will follow the behaviour predicted by the effective WAVG model.
We can now use the local TB corrections for single defects, discussed in the preceding section, to thoroughly investigate the robustness of the proximity effects predicted by the effective model in larger, more disordered systems, such as that in Fig.~\ref{gap_imp}(a). 
Pristine G/MoSe$_2$ and G/WSe$_2$ systems have different topological character, with a normal gap at the Dirac point noted for the former and band inversion for the latter.
Similar to previous studies of the SOC parameter space \cite{Alsharari2016mass,Gmitra2016trivial}, our effective model predicts a semi-metallic state at the boundary of the Mo--like and W--like phases, shown earlier in Fig.~\ref{graphene/TMDC_lattice}(c) when $\chi=0.33$.
The closing and re-opening of the gap in the effective model interpolates between Mo and W parameters and is shown explicitly by the solid green curve in Fig.~\ref{gap_imp}(b).
The gap closure at $\chi=0.33$ exactly coincides with the equivalence in magnitude of the mass and VZ terms, as seen in Fig.~\ref{gap_imp}(c).
The VZ term increases with the concentration of W in the alloyed structure, and beyond the critical composition the system adopts the inverted band characteristics of W--based heterostructures.
To determine the effect of local corrections on this behaviour, we construct a range of $4\times 4$ blocks disordered supercells, as shown in Fig.~\ref{gap_imp}(a), and calculate their band dispersion and spin texture using a tight-binding model which superimposes local atomic--level corrections on top of the constant effective model potential and spin--orbit fields.
The band gaps extracted from these calculations are shown by symbols in Fig.~\ref{gap_imp}(b), together with the \emph{average} VZ and $\Delta$ values across the unit cell in Fig. \ref{gap_imp}(c).
Towards the W--rich side of the plot, the band gaps from the individual TB simulations for composite systems closely follow the effective model prediction.  
The band gaps on the Mo--rich side, however, show a larger variation around the effective model value.
This can be attributed to the more pronounced effects induced by individual W atoms in our model.
Local potentials from W impurities change the overall average mass term throughout the supercell, modulating the band structure and gap.
Since the pristine Mo system is near the transition point, the band structure is more sensitive to perturbations than in the pristine W system.
While the large supercells considered in these calculations allow the contributions from individual impurities to average out to some extent, this effect is smallest for configurations near the pristine limits, which may only have one or two impurities.
We expect to recover the bands and spin textures of the effective model more closely in  realistic--scale, non--periodic systems which will contain a mix of different impurity configurations, even in dilute limits.  
As noted in Sec.~\ref{eff_model}, the magnitude of the mass term extracted from DFT calculations can be sensitive to supercell details. 
To account for this, the (dot-)dashed lines in Fig.~\ref{gap_imp}(b) show the band gap behaviour if the mass term in the effective model is increased (orange) or decreased (blue) by 20\%.
This does not change the qualitative behaviour discussed above, but does shift the critical concentration at which the transition occurs.
Combined with the results for local perturbations, this suggests that the general trend predicted by the effective model, namely gap closing and transition behaviour, are robust against changes in parameters that may occur in realistic systems.

To confirm the phase transition in large mixed systems, we also plot the valley Chern index as a function of $\chi$ in Fig.~\ref{gap_imp} (c). 
Details of this calculation are given in Appendix \ref{top_order}.
Similar to previous reports \cite{Alsharari2016mass,Frank2018Protected}, we find that the band structure in proximitized graphene with VZ SOC give rise to a non-zero Berry curvature in the vicinity of the Dirac points in each valley.  
However, due to time--reversal symmetry the Berry curvatures at the valleys are equal but with a sign change.
This leads to a vanishing total Chern index, but a non-zero valley--projected Chern index, i.e. $C_V=\pm1$\cite{Alsharari2016mass}. 
This index undergoes a sign change when $|\Delta| = |\lambda_{VZ}|$, as shown in  Fig~ \ref{gap_imp} (c)
This is similar to topological effects noted for gapped bilayer graphene\cite{zhang2013valley}, where changing the direction of an interlayer bias similarly changes the sign of the valley Chern index.
This sign change in our system occurs at $\chi=0.33$, further confirming the semi-metallic transition that separates the normal and inverted gap regimes.
As the staggered mass and SOC parameters can be altered by controlling the composition ratio of metallic element, this suggest that the topological behaviour of the graphene/TMDCs can be conveniently tuned.
Since the VZ SOC and the nonzero Berry curvature at each valley in stacked graphene/TMDCs can lead to topologically protected edge states \cite{Frank2018Protected,tiwari2020observation}, we propose experimentally viable alloyed graphene/TMDCs for realization of exotic quantum valley Hall effects.

\section{Conclusions}\label{conclusions}
We have studied the evolution of proximity SOC in a range of different alloyed TMDC layers and their signature on the electronic and spin--orbit characteristics of the graphene layer.
Using the Dirac Hamiltonian and the weighted average SOC parameters, we developed an effective model 
to study the disorder--induced SOC in graphene/TMDCs. 
We found that the effective model based on the disorder concentration can successfully predict the physical behaviour of large composite graphene/TMDC systems.
Furthermore, by controlling the composition ratio of metallic element, one can tune the topological state of alloyed systems.

\begin{acknowledgments}
The authors wish to acknowledge funding from the Irish Research Council under the Laureate awards and the Government of Ireland postdoctoral fellowship program.  
Computing resources used in this study were provided by Trinity Centre for High Performance Computing (TCHPC) and by Irish Centre for High--End Computing (ICHEC).
\end{acknowledgments}


%

\appendix

\section{Methods}
\subsection{Density Functional Theory}\label{DFT}
The electronic dispersion and spin textures of graphene/TMDC heterostructures are calculated using the \textsc{Quantum} ESPRESSO package \cite{Giannozzi_2009}. 
We employ fully relativistic projector augmented wave (PAW) pseudopotentials in combination with the generalized gradient approximation (PBE) for the exchange--correlation potential. 
We set the kinetic energy cutoff to 60 Ry and sample the Brillouin zone with 12 $\times$12 k points. 
We optimize the atomic positions using the quasi-Newton algorithm based on the trust radius procedure. The adjacent layers are bound to each other through the Van der Waals interaction using a semiempirical approach \cite{Grimme2006Semiempirical}. We use the Coulomb truncation \cite{Sohier2017Density} and set a 13\AA~vacuum space to avoid the interactions between spurious images of the sampled supercell along the direction normal to the graphene/TMDC plane.

\subsection{Continuum model}\label{appendixcont}
The continuum model consists of the Dirac Hamiltonian for $p_z$ electrons in graphene ($\mathcal{H}_O$) with additional SOC terms ($\mathcal{H}_{\rm SO}$) that are driven by proximity to TMDC layer, i.e. $\mathcal{H} = \mathcal{H}_O + \mathcal{H}_{\rm SO}$. 
In fact, because the graphene Dirac bands are located in MoSe$_2$ and WSe$_2$ band gap \cite{Gmitra2016trivial}, it is electrons in graphene solely which contribute to the low energy electronics of the heterostructures discussed in this work. 
The orbital, spin-degenerate part of the Hamiltonian is given by 
\begin{equation}\label{orbit}
    \mathcal{H}_O = \hbar v_{\rm F}(\kappa\sigma_x k_x+\sigma_y k_y) + \Delta \, \sigma_z \,,
\end{equation}
where $v_{\rm F}$ is the Fermi velocity, $\sigma_i,\, i=x,y,z$ are the Pauli matrices acting on the orbital space with $\sigma_0$ representing the identity matrix, and $\kappa=+1(-1)$ is the $K (K')$ valley index.
The mass term $\Delta\,\sigma_z$ represents a proximity--induced spin--independent breaking of sublattice symmetry in the graphene layer that opens a band gap at the Dirac points.
The spin-orbit part of the Hamiltonian reads 
\begin{equation}\label{h_soc}
    \mathcal{H}_{SO} = \mathcal{H}_I + \mathcal{H}_{\rm VZ} + \mathcal{H}_{\rm R} + \mathcal{H}_{\rm PIA} + \mathcal{H}_{\Delta {\rm PIA}} \,,
\end{equation}
where
\begin{equation}\label{soc_terms}
\begin{aligned}
	\mathcal{H}_{\rm I} &= \lambda_{\rm I} \kappa \sigma_z s_z, \\
	\mathcal{H}_{\rm VZ} &= \lambda_{\rm VZ} \kappa  s_z, \\
	\mathcal{H}_{\rm R} & =  \lambda_{\rm R} (\kappa\sigma_x s_y-\sigma_y s_x), \\
	\mathcal{H}_{\rm PIA} &= a_G \lambda_{\rm PIA} \sigma_z (k_x s_y - k_y s_x), \\
	\mathcal{H}_{\Delta_{\rm PIA}} &= a_G \Delta_{\rm PIA} (k_x s_y - k_y s_x) \,.
	\end{aligned}
\end{equation}
Here $s_i,\, i=x,y,z$ are the Pauli matrices that operate on the spin space and $a_G$ is the graphene lattice constant.
$\mathcal{H}_{\rm I}$ is the Kane--Mele SOC term which is responsible for spin--independent topological gap. 
The VZ SOC, $\mathcal{H}_{\rm VZ}$, leads to a spin--valley locking shift of the Dirac cone.
The intrinsic and VZ terms can be thought of as the sublattice symmetric and asymmetric contributions to a Kane-Mele type coupling.
This type of SOC does not mix spin channels and is typically very weak in isolated graphene, but can be enhanced by substrate or impurity effects.
The Rashba term, $\mathcal{H}_{\rm R}$, on the other hand, is a substrate--induced SOC which arises due to symmetry--breaking in the direction perpendicular to graphene layer.
The Rashba term acts to couple electrons of different spin orientations introducing in--plane spin textures and spin precession effects in the absence of external magnetic fields.
The $\mathcal{H}_{\rm PIA}$ and $\mathcal{H}_{\Delta {\rm PIA}}$ terms are responsible for renormalization of the Fermi velocity and for spin--dependent band splitting further from the valleys. 
The orbital and spin-orbit parameters of graphene in stacked graphene/TMDC heterostructures can be found by fitting DFT results to the Dirac model Hamiltonian in Eq. \eqref{orbit} -- \eqref{soc_terms}.
The full sets of parameters for pristine and single-impurity configurations are given in Table \ref{Tab:graphene/TMDC_param} and \ref{Tab:impurity_param}, respectively.

\setlength{\tabcolsep}{6pt} 

\begin{table*}[t]
\begin{center}
\begin{tabular}{ccccccccc}
\hline
nr. of W atoms & atom label &  $t$ & $\Delta$ & $\lambda_{\rm R}$  & $\lambda_{\rm I}$ & $\lambda_{\rm VZ}$ & $\lambda_{\rm PIA}$  & $\Delta_{\rm PIA}$ \\
&& [eV] & [meV] &  [meV] & [$\mu$eV] & [meV] & [meV] & [meV]\\ \hline

\hline
1	&	1	&	2.53	&	0.11	&	0.29	&	-1.68	&	0.29	&	-6.3	&	-0.68	\\
	&	2	&	2.53	&	-0.77	&	0.34	&	-10.19	&	0.38	&	-6.31	&	-0.82	\\
	&	3	&	2.531	&	-0.18	&	0.32	&	3.56	&	0.39	&	-6.16	&	-0.92	\\
	&	4	&   2.53    &	-0.38   &	0.33    &	-22.32  &	0.38    &	-6.31   &	-0.82	\\
	&	5	&	2.53	&	0.11	&	0.32	&	14.11	&	0.39	&	-6.31	&	-0.82	\\
	&	6	&	2.53	&	-0.53	&	0.3	&	-143.04	&	0.43	&	-6.31	&	-0.82	\\
	&	7	&	2.53	&	-0.77	&	0.32	&	-3.46	&	0.39	&	-6.31	&	-0.82	\\
    &	8	&	2.53	&	-0.47	&	0.37	&	143.18	&	0.45	&	-6.31	&	-0.82	\\
	&	9	&	2.53	&	-0.51	&	0.32	&	-7.61	&	0.39	&	-6.31	&	-0.82		\\
	\hline
8	&	1	&	2.53	&	-0.46	&	0.51	&	-4.14	&	1.16	&	0.88	&	1.47	\\
	&	2	&	2.531	&	-0.56	&	0.48	&	-8.88	&	1.08	&	-0.45	&	-0.51	\\
	&	3	&	2.531	&	-0.54	&	0.48	&	-0.28	&	1.07	&	-0.45	&	-0.51	\\
	&	4	&	2.531	&	-0.53	&	0.49	&	-10.13	&	1.07	&	-0.45	&	-0.51	\\
	&	5	&	2.531	&	-0.56	&	0.48	&	-11.46	&	1.08	&	-0.45	&	-0.51	\\
	&	6	&	2.531	&	-0.5	&	0.49	&	133.41	&	1.03	&	-0.45	&	-0.51	\\
	&	7	&	2.531	&	-0.56	&	0.49	&	-7.79	&	1.07	&	-0.45	&	-0.51	\\
	&	8	&	2.531	&	-0.65	&	0.45	&	-139.15	&	1.02	&	-0.45	&	-0.51	\\
	&	9	&	2.531	&	-0.55	&	0.48	&	1.16	&	1.07	&	-0.45	&	-0.51	\\
	\hline 
	
 \hline	
 \end{tabular}
\end{center}
\caption{\small Calculated orbital and spin--orbit parameters for composite graphene/TMDC heterostructures with single W and Mo defect. 
The metal atoms in the TMDC layer are labeled in ascending order from one to nine according to their position in the cell. The cell is depicted in the Fig. \ref{SOC_param}(a). The labels run from bottom left to top right corner of the unit cell.
The parameters are calculated by fitting the DFT results to those of continuum model. $t$ is the hopping energy of graphene's $p_z$ electrons, $\Delta$ denotes the sublattice staggered mass, $\lambda_{\rm I}$ the Kane--Mele, and $\lambda_{\rm VZ}$ the valley--Zeeman SOC term. $\lambda_{\rm R}$ is the Rashba spin-orbit coupling, and $\lambda_{\rm PIA}$ and $\Delta_{\rm PIA}$ are the pseudospin--inversion--asymmetry SOC.   }\label{Tab:impurity_param}
\end{table*}

\subsection{Tight Binding}
\label{app:tb}
The Dirac Hamiltonian in Eq. \eqref{orbit} --  \eqref{soc_terms} has an associated TB Hamiltonian which reads \cite{Alsharari2016mass,Kochan2014spin,Gmitra2016trivial}
\begin{eqnarray}\label{Eq:TB}
\mathcal{H} &=&
\sum_{\langle i,j\rangle,\sigma} t~ c_{i\sigma}^\dagger c^{\phantom\dagger}_{j\sigma}+
\sum_{i,\sigma} ~\Delta~ \xi_{c_i}\,c_{i\sigma}^\dagger c^{\phantom\dagger}_{i\sigma} \nonumber \\
&&+\frac{2i}{3}\sum_{\langle i,j\rangle}\sum_{\sigma,\sigma'}c_{i\sigma}^\dagger c^{\phantom\dagger}_{j\sigma'}\left[\lambda_{\rm R}^{ij} \left(\mathbf{\hat{s}}\times \mathbf{d}_{ij}\right)_z\right]_{\sigma\sigma'}\\
&&+\frac{i}{3}\sum_{\langle\langle i,j\rangle\rangle}\sum_{\sigma,\sigma'}c_{i\sigma}^\dagger c^{\phantom\dagger}_{j\sigma'} \left[\frac{\tilde\lambda_{\rm I}^{ij}}{\sqrt{3}}\nu_{ij}\hat{s}_z + 2\tilde\lambda_{\mathrm{PIA}}^{ij} \left(\mathbf{\hat{s}}\times \mathbf{D}_{ij}\right)_z\right]_{\sigma\sigma'}.\nonumber
\end{eqnarray}

Here, the $c_{i\sigma}^\dagger(c_{i\sigma})$ operator creates (annihilates) an electron at atomic site $i$ with spin $\sigma$.
$\xi_{c_i} = \pm 1 $ is a sublattice index and $\hat{s}$ is the spin vector made of  Pauli matrices.
$\mathbf{d}_{ij}$ and $\mathbf{D}_{ij}$ are the unit vectors connecting the nearest neighbor and next nearest neighbors, respectively, and $\nu_{ij}=1(-1)$ defines the trajectory sign, i.e. clockwise (counterclockwise) from  site $j$ to  site $i$. $\tilde \lambda_{\rm I}$ and $\tilde \lambda_{\mathrm{\rm PIA}}$ are the generic sublattice--dependent intrinsic and PIA parameters that in a pristine system with uniform parameters can be written as:
\begin{align*}
	\tilde \lambda_{\rm I}^A &= (\lambda_{\rm I}+\lambda_{\rm VZ}) \\
	\tilde\lambda_{\rm I}^B &= (\lambda_{\rm I}-\lambda_{\rm VZ})\\
	\tilde\lambda_{\rm PIA}^A &= (\lambda_{\rm PIA}+\Delta_{\rm PIA})\\
	\tilde\lambda_{\rm PIA}^B &= (\lambda_{\rm PIA}-\Delta_{\rm PIA})
\end{align*}

At low energies, excellent agreement is found between a TB calculation with uniform parameters and the Dirac Hamiltonian model.  
However, the lattice nature of the tight-binding model allows the strength of various terms to vary spatially, \emph{e.g.} $\lambda_{\rm R} \rightarrow \lambda_{\rm R} (\mathbf{r})$, and for local perturbations to be introduced at specific locations, \emph{e.g.} $\lambda_{\rm R}^{ij} = \lambda_{\rm R} + \delta$.
The variation in parameter strength can be introduced in a number of ways.
David et al consider the direct overlap between orbitals in the two layers using a Slater--Koster approach, and use this to examine how the effective spin--orbit fields vary as a function of twist--angle between the two layers \cite{David2019induced}.
In this work, we use the tight-binding methods to examine how changes to the proximity SOC in the graphene layer, due to alloying the TMDC layer, can be accounted for by locally changing parameters near an impurity.
This is done by comparing the TB and DFT calculations with the same unit cell size and defect locations, and fitting selected perturbation parameters to optimise the agreement between the two. 
Once carefully parameterised, the computationally lightweight nature of TB calculations also allows us to consider much larger systems than ab initio calculations, and investigate the effective proximity effects in realistic, large--scale alloyed graphene/TMDC systems.

\section{Valley Chern number}\label{top_order}
The topological nature of the complex systems can be better verified by evaluation of the valley Chern number. 
The valley Chern index is $C_V=(C_K-C_{K'})/2$ where $C_{K(K')} = \frac{1}{2\pi} \int ~\Omega({\bf k})d\bf{k}$ denotes the Chern number at the non--equivalent Brillouin zone corners. Here the integration runs over momentum space near the $K(K')$ valley and  
\begin{eqnarray}\label{Eq:chern}
    \Omega({\bf k}) = -2 \hbar^2 \sum_n f_n \sum_{n'\neq n} \Im \frac{\langle n{\bf k}|v_x|n'{\bf k}\rangle \langle n'{\bf k}|v_y|n{\bf k} \rangle }{(E_n({\bf k})-E_{n'}({\bf k}))^2}~~~~~~~~
\end{eqnarray}
is the Berry curvature. $n(n')$ is the band index, $f_n$ is the Fermi--Dirac distribution function, and $v_{x(y)}=\partial H/\partial (\hbar k_{x(y)})$ is the velocity along the in--plane direction.

\end{document}